\documentclass[prc]{revtex4}

\usepackage{graphicx,amsmath}
\begin{document}
\title{Interpreting scattering wave functions in the presence of energy-dependent interactions}
\author{Scott Pratt}
\affiliation{Department of Physics and Astronomy,
Michigan State University\\
East Lansing, Michigan 48824-1321}
\date{\today}

\begin{abstract}
In scattering theory, the squared relative wave function $|\phi({\bf q},{\bf r})|^2$ is often interpreted as a weight, due to final-state interactions, describing the probability enhancement for emission with asymptotic relative momentum $q$.  An equivalence relation also links the integral of the squared wave function over all coordinate space to the density of states. This relation, which plays an important role in understanding two-particle correlation phenomenology, is altered for the case where the potential is energy dependent, as is assumed in various forms of reaction theory. Here, the modification to the equivalence relation is derived, and it is shown that the squared wave function should be augmented by a additional factor if it is to represent the emission enhancement for final-state interactions. Examples with relativistic vector interactions, e.g., the Coulomb interaction, are presented.
\end{abstract} 

\maketitle

In many forms of reaction theory the square of the outgoing wave function, $|\phi({\bf q},{\bf r})|^2$, plays the role of a weight, enhancing the probability for emission from coordinate ${\bf r}$ into an asymptotic momentum state ${\bf q}$. One such example is the Koonin equation \cite{koonin,hbtreview}, used for two-particle correlations,
\begin{equation}
\label{eq:koonin}
C({\bf q})=\int d^3r S({\bf r}) |\phi({\bf q},{\bf r})|^2,
\end{equation}
where $q$ is the relative momentum and $S({\bf r})$ is the normalized probability for emitting two particles of the same momentum so that in their center-of-mass frame they are separated by ${\bf r}$. Thus, the squared wave function describes the additional probability for emitting particles due to their final-state interactions. A second method for calculating the probability enhancement is to consider the correction to the relative density of states, which is related to the phase shifts by the Beth-Uhlenbeck formula \cite{bethuhlenbeck,huang,landaulifshitz},
\begin{equation}
\label{eq:bethuhlenbecka}
\Delta\frac{dn}{d\epsilon}=\sum_{\ell}\frac{(2\ell+1)}{\pi}\frac{d\delta_\ell}{d\epsilon}\, ,
\end{equation}
where $\Delta dn/d\epsilon$ is the change of the density of states induced by the potential. The Beth-Uhlenbeck formula is only applicable for volumes $V$ which are much larger than the range of the interaction or the inverse momentum $1/q$. From this perspective, final-state interactions provide an extra weight for emission through the phase shift,
\begin{equation}
\label{eq:Cfromdelta}
C(q)=1+\frac{\sum_\ell [(2\ell+1)/\pi] (d\delta_\ell/dq)}{V q^2/(2\pi^2)}\, ,
\end{equation}
where the denominator is the density of free states and $q$ is the relative momentum. Since the volume $V$ is in the denominator, this relation can be used to experimentally infer the size of the system. For the case of pion-proton correlations, a bump ensues in the correlation function for values of $q$ corresponding to the invariant mass of the delta resonance. In this neighborhood the $\ell=1,I=3/2$ phase shift rises quickly from zero to $\pi$, and the height of the bump provides a quantitative measure of the overall volume. If the scattering particles have intrinsic spin, the denominator in Eq. (\ref{eq:Cfromdelta}) picks up a factor $(2S_1+1)(2S_2+1)$, and the sum over angular momenta in the numerator would be expanded, $\sum_\ell\rightarrow \sum_{\ell,S}(2S+1)$, where $S$ is the total spin. The phase shift would then depend on $S$ in addition to its momentum dependence. Since the inclusion of spin does not alter any of the relations  derived further below, aside from the addition of spin factors, spin will be suppressed throughout the remainder of the manuscript.

It is easy to derive the Beth-Uhlenbeck formula by considering a scattering center inside a large sphere of radius $R$. The partial wave of outgoing momentum $q$ and angular momentum $\ell$ would satisfy the boundary condition, $\sin(qR+\delta_\ell)=0$, which gives the constraint $qR+\delta_\ell=n\pi$. Taking the derivative gives $dn=d\delta_\ell/\pi$, thus proving the relation. The relation can be easily extended to inelastic interactions by considering eigenphases \cite{pratteigenphases}. Eq. (\ref{eq:bethuhlenbeck}) is extraordinarily useful as it also provides the second virial coefficient. Furthermore, since $2d\delta/d\epsilon$ is also the time delay in the scattering of a wave packet, it relates the extra time spent near a scattering center to the change of the density of states, which can be shown to also satisfy the ergodic theorem \cite{danprattergodic} which states that in a thermalized system, a particle populates the neighborhood of a scatterer proportional to the local density of states. Equation (\ref{eq:bethuhlenbeck}) applies even for relativistic motion, or for energy-dependent interactions.

The equivalence between Eq.s (\ref{eq:koonin}) and (\ref{eq:Cfromdelta}) in the large volume limit can be explicitly demonstrated by considering the large volume limit, where the source function in Eq. (\ref{eq:koonin}) can be replaced with $S({\bf r})\approx S({\bf r}=0)$. In that case,
\begin{eqnarray}
\label{eq:clargesource}
C({\bf q})&=&1+S({\bf r}=0)\int d^3r~\left(|\phi({\bf q},{\bf r})|^2-1\right)\\
\nonumber
&=&1+S({\bf r}=0)4\pi\int r^2dr \sum_\ell(2\ell+1)\frac{1}{q^2r^2}\left\{|\phi_\ell(q,r)|^2-|\phi_\ell^{(0)}(q,r)|^2\right\},
\end{eqnarray}
where in the partial wave expansion $\phi_\ell$ are solutions to the one-dimensional radial Schr\"odinger equation, with an asymptotic form $u_\ell(r\rightarrow\infty)=\sin(qr+\ell\pi+\delta_\ell(q))$. To finish demonstrating the equivalence between the Koonin equation and the Beth-Uhlenbeck formula we apply a relation linking the integral of the squared partial waves with the derivatives of the phase shifts \cite{boal},
\begin{equation}
\label{eq:boal}
\int dr \left\{|\phi_\ell(q,r)|^2-|\phi^{(0)}_\ell(q,r)|^2\right\}
=\frac{1}{2}\frac{d\delta_\ell}{dq},
\end{equation}
where $\phi^{(0)}_\ell$ is the partial wave in the absence of the potential. Inserting Eq. (\ref{eq:boal}) into Eq. (\ref{eq:clargesource}),
\begin{eqnarray}
\label{eq:bethuhlenbeck}
C({\bf q})&=&1+S({\bf r}=0)\frac{2\pi}{q^2}\sum_\ell(2\ell+1)\frac{d\delta_\ell}{dq},
\end{eqnarray}
which is equivalent to Eq. (\ref{eq:bethuhlenbecka}) for the case that the source is uniform over a large volume, i.e., $S(r=0)\rightarrow 1/V$.

The phase-shift-based expression of Eq. (\ref{eq:bethuhlenbeck}) is applicable for a large volume, but for small volumes the Koonin form, Eq. (\ref{eq:koonin}), is required. The equivalence between the two equations in the large volume limit makes it clear that one can identify the change of emission probability arising from final-state interactions with either the change in the density of normalized eigenstates, or the change of the amplitude for emitting particles from a specific location into the final state. This equivalence emphasizes the role thermalization plays in justifying the Koonin formula, Eq. (\ref{eq:koonin}), and is also often used to test numerical solutions of the scattered wave function. The general assumption has been that any potential that reproduces the phase shifts, can be used in the Koonin formula, assuming the range of the potential is much smaller than the size of the source being explored. However, as will be shown below, this equivalence becomes invalid when the potential becomes energy dependent.

The equivalence relation for partial waves, Eq. (\ref{eq:boal}), can be derived from the Schr\"odinger equation, assuming an energy-independent potential, and using a similar approach as to what is used to derive the basic relations of effective range theory \cite{prestonbhaduri,bethe}. The goal of this brief paper is to show how the equivalence relation is modified for the case of energy-dependent potentials. The modification will involve augmenting the squared wave function by a simple multiplicative factor that depends on the derivative of the potential with respect to $q$. Relativistic motion of a particle interacting with a vector potential, which has an effectively energy-dependent interaction when mapped to the Schr\"odinger equation, will be considered in detail. After deriving the modification to Eq. (\ref{eq:boal}) below, the classical limit will be considered, where it will be shown that the same correction factor arises from considering the probabilistic enhancement classically, in the presence of an energy-dependent potential.

The derivation begins by considering the following form for the Schr\"odinger equation,
\begin{equation}
\label{eq:schrodinger}
\left\{-\partial_r^2+\frac{\ell(\ell+1)}{r^2}+U(q,r)\right\}\phi_\ell(r)
=q^2\phi_\ell(r).
\end{equation}
Here, $U(q,r)$ would equal $2\mu V(r)$ for the non-relativistic case with no energy-dependent interactions, and $\mu$ would represent the reduced mass. We use the term {\it energy-dependent interaction} to clarify that the $q$ dependence in $U(q,r)$ is a function of the asymptotic kinetic energy and the position. The phrase {\it momentum-dependent interaction} sometimes refers to exactly such interactions. For instance, in \cite{myersswiatecki,bandyopadhyay} the nuclear optical potential is expressed as a function of the magnitude of the local momentum as calculated from classical arguments, which means that effectively $U(q,r)$ can be considered energy-dependent. If the potential were expressed as a function of $r$ and gradients, i.e., $U(\nabla,r)$, it would not satisfy our definition of being energy-dependent.

Considering the solution to Eq. (\ref{eq:schrodinger}) for two neighboring values of the asymptotic momentum $q$ and $q'$,
\begin{eqnarray}
&&\int_0^R dr~\left[-\phi'(q',r)\partial_r^2\phi(q,r)
+\partial_r^2\phi'(q',r)\phi(q,r)\right]\\
\nonumber
&&\hspace*{40pt}+\int_0^R dr~ \left[U(q,r)-U(q',r)\right] \phi'(q',r)\phi(q,r)
=(q^2-q'^2)\int_0^R dr~ \phi'(q',r)\phi(q,r),
\end{eqnarray}
integration by parts combined with keeping only terms linear in $q-q'$ yields
\begin{equation}
\left[-\phi'(q',r) \partial_r\phi(q,r)+\partial_r\phi'(q',r)\phi(q,r)\right]_{r=R}
=(q-q')\int_0^R dr~ \phi^2(q,r) \left(2q -\frac{\partial}{\partial q}U(q,r)\right).
\end{equation}
Assuming $R$ is sufficiently large to justify use of the asymptotic form of the wave function, $\phi_\ell(q,r)\sim \sin(qr+\ell\pi+\delta_\ell(q))$, and after substituting $\phi(q,r)=\phi(q',r)+(q-q')\partial_q \phi(q,r)$,
\begin{equation}
\frac{1}{2}\left(R+ \frac{d\delta}{dq}\right)=\int_0^R dr~ \phi^2(q,r) \left[1 -(1/2q)\partial_qU(q,r)\right].
\end{equation}
After subtracting the same quantity with zero potential (and thus zero phase shift), one finds the generalized form of Eq. (\ref{eq:boal})  relating $d\delta/dq$ and the wave function,
\begin{equation}
\label{eq:generalresult}
\int_0^R dr~ \left\{|\phi(q,r)|^2 \left[1 -(1/2q)\partial_qU(q,r)\right]
-|\phi^{(0)}(q,r)|^2\right\}=\frac{1}{2}\frac{d\delta}{dq}\,.
\end{equation}
The factor $[1 -(1/2q)\partial_qU(q,r)]$ is the same as found in \cite{schlomo} when calculating the local density of states in a Green's function approach. The expression can be summed over partial waves to find the analogous expression for plane waves, where $\phi({\bf q},{\bf r})|^2$ will also be augmented by the factor $[1 -(1/2q)\partial_qU(q,r)]$. However, the extension to the outgoing plane-wave case relies on the assumption that $U(q,r)$ is independent of $\ell$. This assumption is often violated in scattering phenomenology for nuclear physics, where different forms of the scattering potential are applied for different $\ell$.

One example where a $q$ dependence for $U$ arises naturally is the Klein-Gordon equation for a vector potential $V$,
\begin{equation}
[E-V(r)]^2\phi=(-\nabla^2+m^2)\phi.
\end{equation}
After setting $q^2=E^2-m^2$, the Klein-Gordon equation can be equated to the Schr\"odinger equation, Eq. (\ref{eq:schrodinger}), with the potential $U$,
\begin{equation}
U(q,r)=2EV(r)-V^2(r),
\end{equation}
which, since $\partial_q E=q/E$, is momentum dependent so the constraint of Eq. (\ref{eq:generalresult}) becomes
\begin{equation}
\int_0^R dr~ \left[|\phi_\ell(q,r)|^2 \left(1 -V(r)/E\right)-|\phi_\ell^{(0)}(q,r)|^2
\right]=\frac{1}{2}\frac{d\delta_\ell}{dq}\,.
\end{equation}
Thus, if one uses $|\phi|^2(1-V/E)$, rather than $|\phi|^2$, to describe the probability enhancement, it will be consistent with the correction to the level density. This result should not be surprising, since the density for a scalar field with a gauge field, $V=eA_0$, behaves as $\phi(i\partial_t-eA_0)\phi$, and for an eigenstate becomes $|\phi|^2(E-V)$.   This explicit appearance of the vector potential into physical quantities related to the field is characteristic of gauge invariance, which requires derivatives to be modified, $\partial_\mu\rightarrow\partial_\mu+ieA_\mu$.

The result is somewhat different when one considers the relative motion of two particles of finite mass, $m_1$ and $m_2$. Although the following expression neglects retardation effects, the analogy of the Klein-Gordon equation for the relative wave function can be written as
\begin{equation}
[E-V(r)]\phi=\sqrt{(-\nabla^2+m_1^2)}\phi+\sqrt{(-\nabla^2+m_2^2)}\phi\, .
\end{equation}
Solving for $\nabla^2\phi$ allows one to write an energy-dependent Schr\"odinger equation for $\phi$,
\begin{eqnarray}
-\nabla^2\phi&=&q^2\phi+U(q,r)\phi\,,\\
\nonumber
U(q,r)&=&q^2-\frac{1}{4}\left\{(E-V(r))^2-2(m_1^2+m_2^2)
+\frac{(m_1^2-m_2^2)^2}{(E-V)^2}\right\}\, ,\\
\nonumber
q^2&=&\frac{1}{4}\left\{E^2-2(m_1^2+m_2^2)+\frac{(m_1^2-m_2^2)^2}{E^2}\right\}\, .
\end{eqnarray}
For the limit that $V<<E$, one can keep only the first order terms in $V$, and $U$ can be written as $2\mu e^2/r$ where the effective mass $\mu$ is
\begin{equation}
\label{eq:mu}
\mu=\frac{E}{4}-\frac{(m_1^2-m_2^2)^2}{4E^3}.
\end{equation}
In the non-relativistic limit, $E=m_1+m_2$, and one recovers the usual expression for the effective mass, $\mu=m_1m_1/(m_1+m_2)$, which has no momentum dependence. 

The effects of strong interactions can be described by phase-shifts even when Coulomb interactions are present, and as can be seen below, an equivalence relation can be derived linking the derivative of the phase shift to the integral of $|\phi(q,r)|^2-|\phi_0(q,r)|^2$, where $\phi_0$ includes the Coulomb interaction. Assuming the Coulomb interaction is small enough to justify using the approximation $U_{\rm Coul}=2\mu(q)e^2/r$, the asymptotic wave function becomes
\begin{eqnarray}
\phi_{\ell}(q,r)&\sim& \sin(qr+\eta\ln(r)+\delta_\ell(q))\, ,\\
\nonumber
\eta&=&\mu e^2/q.
\end{eqnarray}
One can repeat the steps used to derive Eq. (\ref{eq:generalresult}), but keeping in mind that $\phi^{(0)}$ was solved with the Coulomb interaction. This results in the relation
\begin{equation}
\label{eq:coulombresult}
\int_0^R dr~ \left\{|\phi_\ell(q,r)|^2 \left[1 -(1/2q)\partial_qU_s(q,r)-(e^2/qr)\partial_q\mu \right]
-|\phi_\ell^{(0)}(q,r)|^2\left[1 - (e^2/qr)\partial_q\mu\right]
\right\}=\frac{1}{2}\frac{d\delta_\ell}{dq}\,.
\end{equation}
Thus, even if the short-range potential $U_s(q,r)$ has no momentum dependence, the Coulomb interaction modifies the equivalence relation by the incorporation of the simple factor $[1 - (e^2/qr)\partial_q\mu]$, which has a momentum dependence due to the fact that $\mu$ depends on $q$ once the motion becomes relativistic, as shown in Eq. (\ref{eq:mu}).

The modification factor $[1- (e^2/qr)\partial_q\mu]$ also appears when one considers the classical analog of the wave function. Classically, particles of outgoing momentum ${\bf q}$ are enhanced by the factor describing the focusing of phase space \cite{coulomb_kim,coulomb_dan}. In these previous studies the role of the squared wave function was shown to be
\begin{equation}
|\phi({\bf q},{\bf r})|^2\rightarrow \frac{d^3q_0}{d^3q},
\end{equation}
where $q_0$ is the momentum at the initial separation ${\bf r}$. The equivalence works whenever $qr\gg 1$.  Non-relativistically, energy conservation implies
\begin{equation}
q^2=q_0^2+2\mu\frac{e^2}{r}=q_0^2+U(q,r)\,.
\end{equation}
If one ignores the $q$-dependence of $U$, $q_0dq_0$ will equal $qdq$. Indeed, this independence was assumed when solving for the wave function at fixed $q$, so the classical-quantum equivalence can be stated as 
\begin{equation}
\left\langle|\phi({\bf q},{\bf r})|^2\right\rangle\rightarrow \frac{q_0}{q}=\sqrt{1-\frac{2\mu e^2}{q^2r}}\, ,
\end{equation}
where the $\langle...\rangle$ denotes an average over angles so that $d^3q=4\pi q^2dq$. This will be identical to the squared wave function if one solves the Schr\"odinger equation for a fixed mass $\mu(q)$ in the limit $qr\gg 1$. However, the factor $q_0dq_0/qdq\ne 1$ if $U$ depends on $q$. In that case differentiating the energy-conservation relation gives,
\begin{equation}
\frac{q_0dq_0}{qdq}=1-(1/2q)\partial_q U(q,r)\, ,
\end{equation}
which is the identical factor used to modify the equivalence relation in Eq. (\ref{eq:generalresult}). This emphasizes the physical origin of the modification factor. It also suggests that for most applications involving the application of relative wave functions for relativistic $q$, one should incorporate the factor $[1-(1/2q)\partial_qU]$ into $|\phi|^2$ if it is to be interpreted as a probability enhancement.

The Coulomb modification factor above only comes into play for relativistic motion where $q$ is not much smaller than the total energy. In the limit of large $q$, $qr$ is also large and one can justify the classical expression. Considering the equal mass case of Eq. (\ref{eq:mu}), where $\mu=E/4$, the large-$q$ limit of the phase space focusing factor becomes
\begin{equation}
\label{eq:coulombtail}
\frac{d^3q_0}{d^3q}\approx 1-\frac{\mu e^2}{q^2r}-\frac{e^2}{\mu r}\, .
\end{equation}
The last term, which arises from the $q$ dependence of $\mu$, is negligible for small $q$ where the motion is non-relativistic, but is four times larger than the second term in the relativistic limit. Also of note, both terms fall as $1/q$ asymptotically, in contrast to the $1/q^2$ behavior of the non-relativistic expression.

For most applications, the energy-dependence of the effective mass can be ignored. This is certainly the case for low-energy reactions. It is also the case for the vast majority of applications involving two-particle correlations for high-energy collisions, since most analyses focus on correlations at small relative momentum. At large relative momentum, corrections for the $q$ dependence of the potential become important, but that is also the region where interactions matter less due to the large competing phase space associated with free motion. For instance, these effects could be important for modeling $\pi\pi$ interactions in the region of the $\rho$ meson  \cite{rhopeak_rapp,rhopeak_bauer}, where the decaying pions are highly relativistic. However, for sources much larger than one femtometer, $qr\gg 1$, one can model observables using statistical arguments based on derivatives of the phase shifts, and avoid applying $|\phi|^2$ weights. Another class of observables where these effects might be important are analyses of charge balance functions \cite{balance_bass,balance_cheng}, which involve like-sign subtractions using many pairs, especially in a high-energy heavy ion collision. In this case, although the effect of Coulomb interactions at large $q$ is small on a pair-by-pair basis, Coulomb effects are magnified by the large number of pairs in a high multiplicity event, so that the more complicated structure of Eq. (\ref{eq:coulombtail}) is required. Energy dependent interactions are also applied to other forms of reaction theory, e.g., the optical models of \cite{myersswiatecki,bandyopadhyay}. Since both the phase shifts and modified density are also of interest in many of these applications, the modification factor could also be relevant. In these cases the energy dependence is not a relativistic effect, and the characteristic momentum scale for the energy dependence is not the rest mass, but instead is determined by some other phenomenological scales. Thus, the importance of the modification factor must be assessed on a case-by-case basis.

\section*{Acknowledgments}
Discussions with Jerry Miller, Pawel Danielewicz, Shalom Shlomo and Li Yang are gratefully acknowledged. Support was provided by the U.S. Department of Energy, Grant No. DE-FG02-03ER41259.

\end{document}